# Heavy doped ABA stacked trilayer graphene: triple splitting of its Raman G peak

By *S. S. Lin,* [1,2,3] [*] *B. G. Chen,* [3] *W. Xiong,* [3] *Y. Yang,* [3] *L. M. Tong,* [3] *Y. Xu,* [1] [*] *J. Luo* [1]

[*] Dr. S. S. Lin, Dr. B. G. Chen, Prof. L. M. Tong, Y. Xu, J. Luo
1, Department of Information Science and Electronic Engineering, Zhejiang University, Hangzhou 310027, (P. R. China).
2, Manchester Centre for Mesoscience and Nanotechnology, School of Physics and Astronomy, University of Manchester, Manchester M13 PL, UK
3, State Key Laboratory of Modern Optical Instrumentation, Department of Optical Engineering, Zhejiang University, Hangzhou, 310027, (P. R. China)
E-mail: (shisheng.daniel.lin@gmail.com (SS LIN) yangxuuiuc@gmail.com (X Yang))

Keywords: (Graphene, doping, Raman, G peak, triple-splitting)

Monolayer graphene (MLG) and few-layer graphene (FLG) are giant flat moleculars with a few extraordinary properties induced by relative quantum mechanics decription of their two dimensional carriers gases, such as micro-scale ballistic transport,[1] room temperature abnormal quantum Hall effect,[2] fine-structure constant related absorption coefficient.[3] Besides, the tunable gate doping promises graphene as a good candidate of replacing of silicon in the modern electronic industry. Graphene also has outstanding thermal conducitivity[4] and high mechanical strength.[5] All these properties allows graphene research continues to be a hot point for expoliting interesting opto-electronic applications.[6-7]

However, the zero band gap of MLG severly limits its otherwise broader optoelectronic applications. Thus, scientists keep exploring other advanced ultrathin 2D semiconductor which could possess a band gap suitalbe for the applictions in the area of light-emitting diodes (LEDs), solar cells and high on/off ratio transistors.[8-10] Strikingly, the band gap of bilayer graphene (BLG) and trilayer graphene (TLG) can be tuned under an electric field normal to the graphene plane.[11] As the electron hopping among the layers is permitted, FLG presents an electronic structure very different from MLG. The band gap of AB stacked BLG can be open up to 250 meV under an asymmetric gate effect.[11] Moreover, a flourished interest in TLG is triggered due to a tunable band gap.[12-15] There are two distinct choices of forming TLG: one is ABA stacked and the other is ABC stacked. The physical properties of ABA stacked TLG are different from ABC stacked TLG due to the distinct interlayer coupling. [16-17] The gap between conduction band and valence band can be open up to 120 meV in ABC stacked TLG while more overlaped in ABA stacked TLG under an electric field normal to the graphene planes. [18] ABA stacked TLG has a unique electronic structure consisting of both massless MLG-like and massive bilayer graphene-like Dirac subbands and, consequently, exhibits many exciting phenomena, such as Landau level crossing in quantum hall effect. [19]

Understanding the phonon decays from a microscopic point of view, in particular, based on the electron phonon coupling (EPC), is a key step to improve or design the transport and optical properties of graphene. [1, 20-24] Furthermore, the EPC also plays a significant role in many phenomena such as the quasiparticle dynamics[25] and potential superconductivity in graphene system.[26-27] EPC strongly influences the phonon frequencies and gives rise to kinks in the phonon dispersion around Γ and K points, named as Kohn anomalies. [28-30] Raman spectra is a powerful technique to detect the vibrational, electronic and EPC properties of graphene. [31-32] The Raman G peak in graphene corresponds to the doubly degenerate $E_{2g}$ phonon at Γ point. Under this context, the position and full width at half maximum (FWHM) of G peak are important fringprints of the EPC in MLG. [33-34] Raman spectra are also used to resolve the layer numbers of graphene as the interlayer coulping is varied by interaction among graphene planes. [32] From MLG to BLG to TLG, the G peak positions increase gradually. Moreover, theoritical work on TLG reveals that G peak can be split into two and three in BLG and TLG, respectively. [35,36] Double-G peak splitting in BLG has been observed in electric field and unintentional doped BLG system due to the broken of inversion symmetry.[37-39] Nevertheless, since the pioneering work by Ferrari et al, [32] no triple G peak splitting has been experimentally observed to the best of our knowledge. Herein, we show that clear triple G peak splitting can be achieved in unintentional doped and nitrogen doped ABA stacked TLG. We propose two physical mechanisms to account for the obvious splitting behavior: one is the different EPC strength for the three components of G peak in ABA stacked TLG and the other is varied interaction strength among layers in ABA stacked TLG caused by doping.

Microcleavage method was used to exfoliate graphene on 300 nm $SiO_2$/Si substrate. [39] Figure 1a shows the optical microscope image of graphene, which consists of MLG, BLG and TLG. The constrast of the MLG, BLG and TLG increases linearly as a function of layer number due to the increased optical absorption (Figure 1b). [3] The 2D band of TLG can be deconvoluted into six peaks, which is a typical characteristic of TLG.[13] Figure 1c shows that the intensity ratio of I(G)/I(Si) (I(Si) points to the inensity of optical phonon around 520 $cm^{-1}$ of silicon) increases linearly as a function of layer numbers, which confirms TLG has been obtained.[40] The Raman 2D mode is expected to be affected by the elctronic properties since it arises from a double resonance process that involves transitions among various electronic states. Lui et al. reported imaging stacking order in TLG using distinctive Raman 2D band and consistently observed an enhanced shoulder in the 2D band of ABC stacked TLG.[13] While the 2D band of the TLG herein does not exhibit such feature, it is in a ABA stacked form. Figure 1d shows a typical field effect transistor (FET) device based on TLG, where electrodes shape are defined by photolithography and 5-nm Ni/50 nm-Au are deposited through e-beam evaporation as electrodes materials.

Figure 2 shows the transfer characteristic of the FET based on TLG and the source-drain current ($I_{ds}$) has a strong dependence on the gate voltage. However, the Dirac point can not be observed up to 200 V. Thus, at zero gate voltage, the carrier density should exceed $2 \times 10^{13}$ cm$^{-2}$.[39] The inset shows that the Raman G peak has been split into three peaks, denoted as G1, G2 and G3 peak, respectively. Recently, there is a few reports that G peak of TLG can be split into two peaks under sulfuric acid doping, [41] NO$_2$ doping [42] or FeCl$_3$ intercalated.[43] However, most of the G peak of TLG reported in literatures presents an symmetric line shape and no triple splitting has been observed.

Figure 3a-c show the schematic illustration of atomic vibration modes of G1, G2 and G3 corresponding to three optical phonons: $E_a$□, $E$□ □and $E_b$□, respectively. Figure 3d shows the G peak can be fitted by three Lorenz peaks, denoted as G1, G2, G3 peak, respectively. In contrast, Figure 3e shows the symmetry G peak observed

most often in another TLG in agreement with that in Ref. 29. [32]As the TLG is unintentionally doped according to FET characterization (Figure 2),[39] it is reasonable to deduce that the G peak splitting is induced by asymmetric impurity doping of the top and bottom layer of TLG. We have measured the Raman G peak spectra for the nanoripple on the TLG, we found that the G peak is symmetric (Figure S1) and thus it could be inferred that the G peak splitting is not induced by residual strain in the TLG.

Within second-order perturbation theory, the phonon self energy can be approximated as: [44]

$$\Pi(\omega_q, E_F) = 2 \sum_{kk'} \sum_{ss'} \sum_{jj'} \frac{\left|g^{v}_{(k+q)j',kj}\right|^2}{\hbar w_q^{s,s'} - E_{kk'}^{eh} + i\gamma_q/2} \times (f_h - f_e) \quad (1)$$

Where $\hbar$ is the planck constant; k and k' are wave-vectors for the initial and final electronic states, respectively; s denotes conduction band and s' denotes valence band; j and j' are then band indices; $q = k - k'$ is the phonon wave-vector; $E_{kk'}^{eh} = (E_{k'}^{e} - E_{k}^{h})$ is the e-h pair energy; $w_q^{s,s'}$ is the phonon frequency; $\gamma_q$ is the phonon decay width; $f_h$ and $f_e$ are the Fermi distributions for holes and electrons, respectively; $g^{v}_{(K+q)j',kj}$ is the EPC matrix element, which is defined as:

$$g^{v}_{k'j',kj} = \sqrt{\frac{\hbar}{2Mw_q}} <k',s',j'\left|\frac{\delta V_{scf}}{\delta u_q}\right|k,s,j> \quad (2)$$

Where M is the carbon atom mass, $\delta V_{scf} = V_{scf}(u_q^v) - V_{scf}(0)$ is the variation in the self-consistent potential field due to the perturbation of a phonon with wave vector q, $|k,s,j>$ is the electronic Bloch state.

From Eq. 1, only when $\hbar w_q > E_{kk'}^{eh}$, the quantity $\hbar w_q - \hbar w_q^0 = \text{Re}(\Pi(w_q, E_F))$ is positive and it is negative when $\hbar w_q < E_{kk'}^{eh}$. If $\text{Re}(\Pi(w_q, E_F)) > 0$, a phonon hardening in $w_q$ occurs while a phonon softening occurs if $\text{Re}(\Pi(w_q, E_F)) < 0$. The distribution of positions of G1, G2, G3 peaks is presented in Figure 3a. The G1, G2, G3 peak positions fluctuate around 1578 cm$^{-1}$, 1588 cm$^{-1}$ and 1590 cm$^{-1}$, respectively. Theoretical calculations suggest the G peak can be split into three peaks locating at 1586 cm$^{-1}$, 1588 cm$^{-1}$ and 1593 cm$^{-1}$, respectively.[35-36] We note that G1 peak of the experimental value (1578 cm$^{-1}$) redshifts compared to the theoretical value (1586 cm$^{-1}$). The softenning of $E_a$ phonon should be caused by the heavy doping of TLG, which results in $\hbar w_q < E_{kk'}^{eh}$ and corresponding the soften (redshift) of G1 peak.[36] Using first principle method, Yan et al. calculated $g_{(K+q)j',kj}^{v}$ for $E_a$, $E_b$ and $E$ mode in TLG and concluded $g_{(K+q)j',kj}^{v}$ varies for $E_a$, $E_b$ and $E$ mode (the EPC strength for $E_b$ is much smaller than $E_a$) in TLG due to different selection rule.[36] The large value of $g_{(K+q)j',kj}^{v}$ for $E_a$ phonon is in consistent with the large red shift of G1 peak compared with the theoritical value. As the EPC strength for $E_b$ is smallest among all, no obvious G3 peak shift will be observed, which is in well accordance with the experimental results.

Figure 4 shows the band structure of ABA stacked TLG, where the transitions contributing $E$ mode and $E$ mode are represented by solid red and dotted blue lines, respectively. A shift of $E_F$ modifies the type and number of transitions contributing to $\Pi(\omega_q, E_F)$. Both symmetry and energy allowed transitions of V$_2$-C$_2$ and V$_1$-C$_1$ still contribute to $E_a$ phonon when $E_F$ is shifting away from the Driac point. However, compared to $V_2$-$C_2$ and $V_1$-$C_1$ trainsitions, interband transitions of $V_1$-$C_2$ or $V_2$-$C_1$ need much higher energy. As Fermi level shift down into the valence band, the $V_2$-$C_2$ and $V_1$-$C_1$ trainsitions become a virtual process, which will not cause the shift of G2 peak to a large extent. This agrees well with the almost unchanged phonon energy of G2 peak compared with the theoretical predictions.

Figure 5a shows the distribution of G peak as function of sample location in a large TLG. The G1, G2 and G3 peak positions scatter in a range and are not a constant. This can not solely explained by the different EPC effect mentioned above. Thus, there must be another intrinsic process which influences the G peak positions. In fact, in addition to the EPC effect, doping can also alters the interlayer interaction strength ($\varepsilon$) and results in the fluctuation of G peak position. The vibration mode can be calculated through tight-binding method applying on the two inequivalent atoms in graphene layers of the TLG. Assuming the interaction strength between the top layer and middle layer is $\varepsilon_1$ and the middle layer and bottom layer is $\varepsilon_2$, the reduced Hamiltonian for TLG can be expressed as:[35]

$$H = \begin{pmatrix} E_0 & \varepsilon_1 & 0 \\ \varepsilon_1 & E_0 + \delta & \varepsilon_2 \\ 0 & \varepsilon_2 & E_0 \end{pmatrix}$$

Where the $\sigma$ is accounting for the change of the on-site energy in the middle layer of trilayer graphene. Solving the secular equation $\det(H - \lambda I) = 0$, one can obtain the eigenvalues are $\lambda_1 = E_0 + \delta/2 - \sqrt{(\delta^2 + 4\varepsilon_1^2 + 4\varepsilon_2^2)}/2$, $\lambda_2 = E_0$, $\lambda_3 = E_0 + \delta/2 + \sqrt{(\delta^2 + 4\varepsilon_1^2 + 4\varepsilon_2^2)}/2$. Through this relationship, we can have:

$$\lambda_{32} = \lambda_3 - \lambda_2 = \delta/2 + \sqrt{(\delta^2 + 4\varepsilon_1^2 + 4\varepsilon_2^2)}/2,$$

$$\lambda_{12} = \lambda_1 - \lambda_2 = \delta/2 - \sqrt{(\delta^2 + 4\varepsilon_1^2 + 4\varepsilon_2^2)}/2,$$

$$\lambda_{31} = \lambda_3 - \lambda_1 = \sqrt{(\delta^2 + 4\varepsilon_1^2 + 4\varepsilon_2^2)}, \text{ Then: } \delta = \lambda_{12} + \lambda_{32}, \ \varepsilon_1^2 + \varepsilon_2^2 = (\lambda_{31}^2 - \delta^2)/4.$$

Figure 5b shows the calculated value of $\varepsilon_1^2 + \varepsilon_2^2$ as a function of sample location at a series value of $\delta$. $\varepsilon_1^2 + \varepsilon_2^2$ ranges from 23 to 37 and has only a slight dependence on the value of $\delta$ (2.5 to 3.5), however, varies from location to location in a broad range. This means that the flucatution of G peak position is related with the varialbe $\varepsilon$ in the TLG (More G peak position analyses can be found in Figure S2), which should be related with spatially varied unintential doping level on a microscopic level. Compared with isolated MLG with no interlayer tunneling, the carrier hopping between the different layers in TLG is relatively strong and thus the carrier density distribution has a strong influence on the $\varepsilon$. A similiar case has been observed in BLG by scanning tunnelling microscopy measurements.[45]

In summary, for the first time, we have observed the obvious triple G peak splitting of TLG. The G peak splitting can be quantatively understood through the different EPC strength of $E_a$, $E_b$ and $E_a$ modes. In addition, the fluctuation of G peak can also be understood from the view of the varied interaction strength among graphene layers of TLG, which is induced by nonuniform hole doping at the microscopic level.

*Experimental section*

Microcleavage method was used to exfoliate graphene on 300 nm $SiO_2$/Si substrate. A field effect transistor (FET) device based on a TLG was fabricated through photolithography and e-beam evaporation techniques. Good ohmic contacts (The inset in Figure 2) between the source/drain electrodes and BLG have been achieved by depositing 5nm Ni and 50nm Au subsequently on BLG. We have carried out electrical measurements this FET and the transfer characteristics (Figure 2) show it is p-type (The Dirac point is a positive value larger than 150V, corresponding to a hole concentration larger than $1.1 \times 10^{13}$ cm$^{-2}$) . The performances of TLG FETs were evaluated using a Keithley 4200 semiconductor analyzer at room temperature. The Raman studies were carried out with a Renishaw micro-Raman spectrometer at 514 nm excitation wavelength.

*Acknowledgements*

S. S. LIN should give great thanks to Andre K Geim for the fellowship supporting his research in the University of Manchester. S. S. LIN also thanks to funding support from State Key Laboratory of Modern Optical Instrumentation (111306*A61001) and the China postdoctoral science foundation (20100480083 and 201104714)

**Figure 1.** (a) Optical microscope image of MLG, BLG and TLG. (b) Typical Raman spectrum of the TLG, which can be deconvoluted into six Lorenz peaks. (c) The intensity ratio of G peak versus Si peak as a function of layer numbers. (d) A typical optical microscope view of a FET device based on TLG.

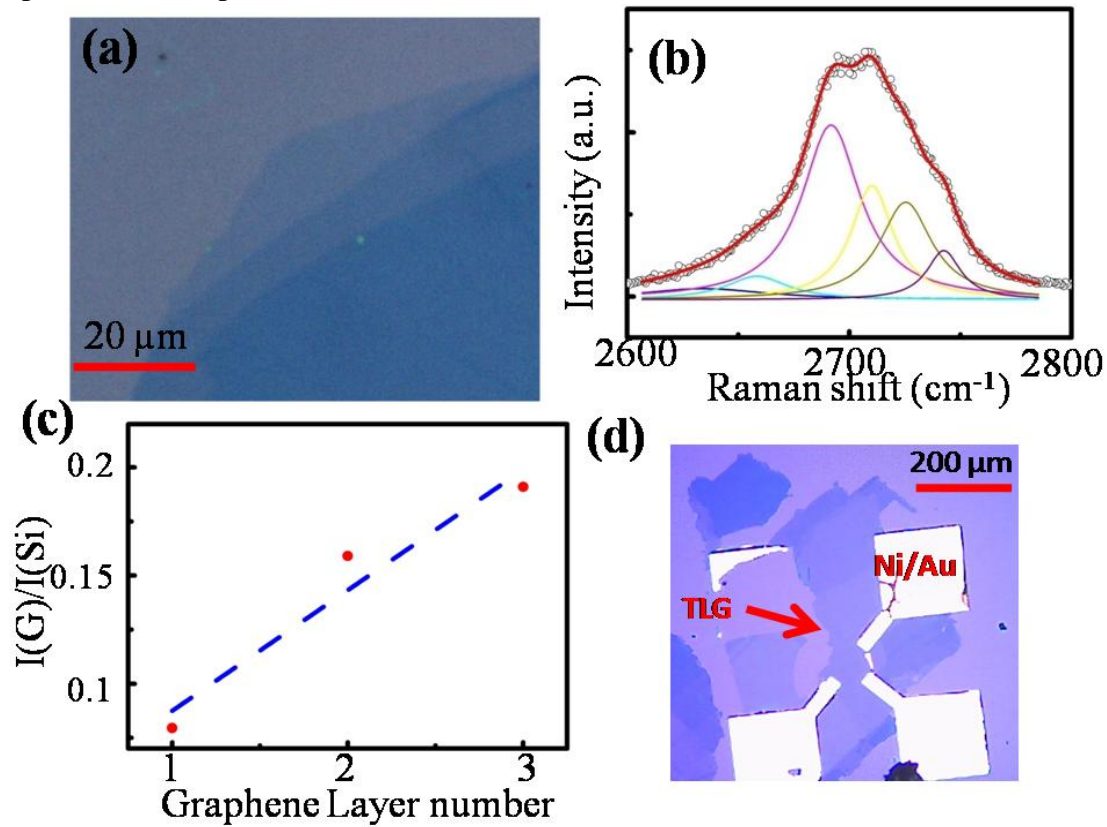

**Figure 2.** Transfer characteristics (dependence of source-drain current ($I_{ds}$) on the gate voltage) of a typical hevey doped TLG, which shows that the Dirac point is locating at a value larger than 150 V. The right-up inset shows the dependence of source-drain current on the source-drain voltage, which indicates that a good ohmic contact between the Ni-Au alloy between graphene has been formed. The left-down inset shows a typical Raman spectrum of this hevey doped TLG, which can deconvolted into three Lorenz peaks.

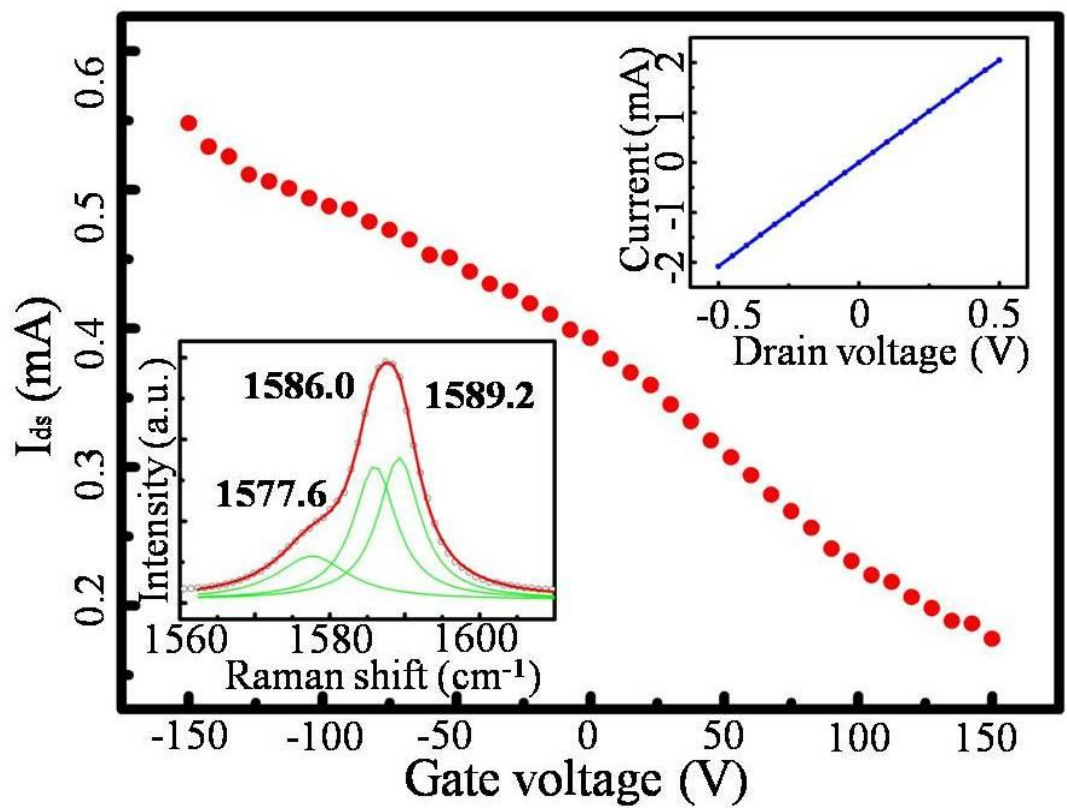

**Figure 3.** (a)-(c) Atomic displacement of the three split optical branches: G1, G2 and G3, corresponding to $E_a$, $E$ and $E_b$, respectively. (d) Typical Raman spectrum of a unsymmetric G peak, which can deconvoluted into three peaks. (e) Typical Raman spectrum of a symmetric G peak.

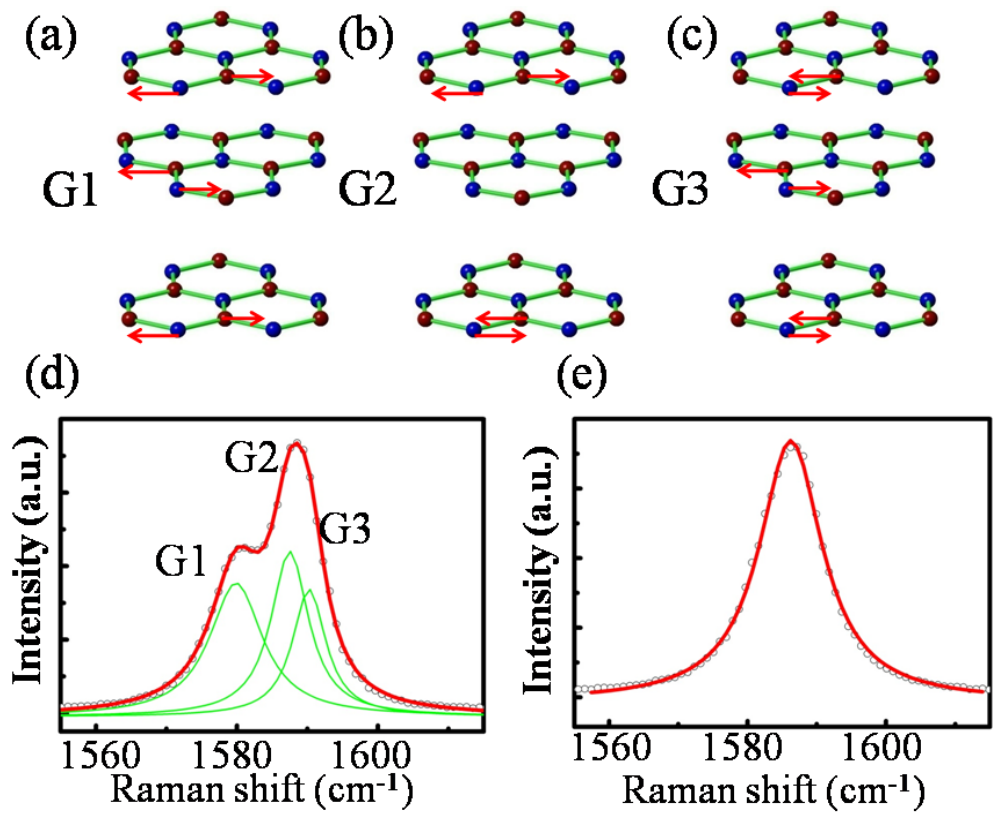

**Figure 4.** Schametic illustration for electronic transitions contributing to E′ mode and E″ mode at different Fermi level for TLG (a) $E_F$ level passes Dirac point. (b) $E_F$ is far away from Dirac point.

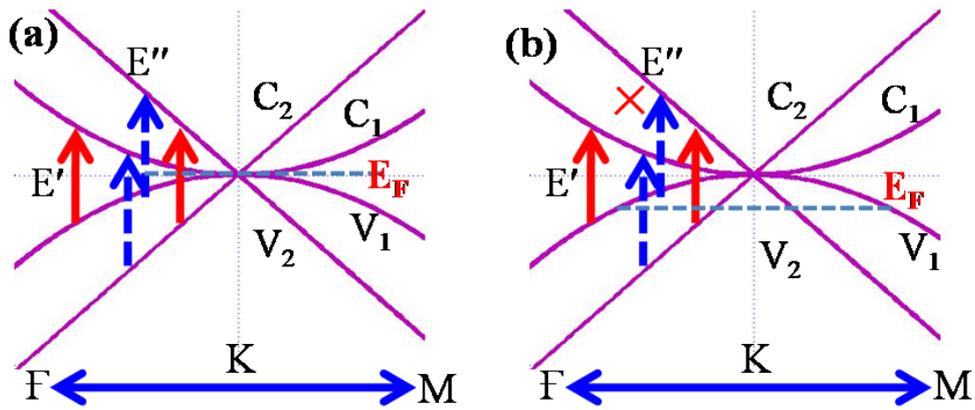

**Figure 5.** (a) Variation of G1, G2 and G3 peak positions as a function of sample locations. (b) the variation of $\varepsilon_1^2 + \varepsilon_2^2$ at different value of $\delta$.

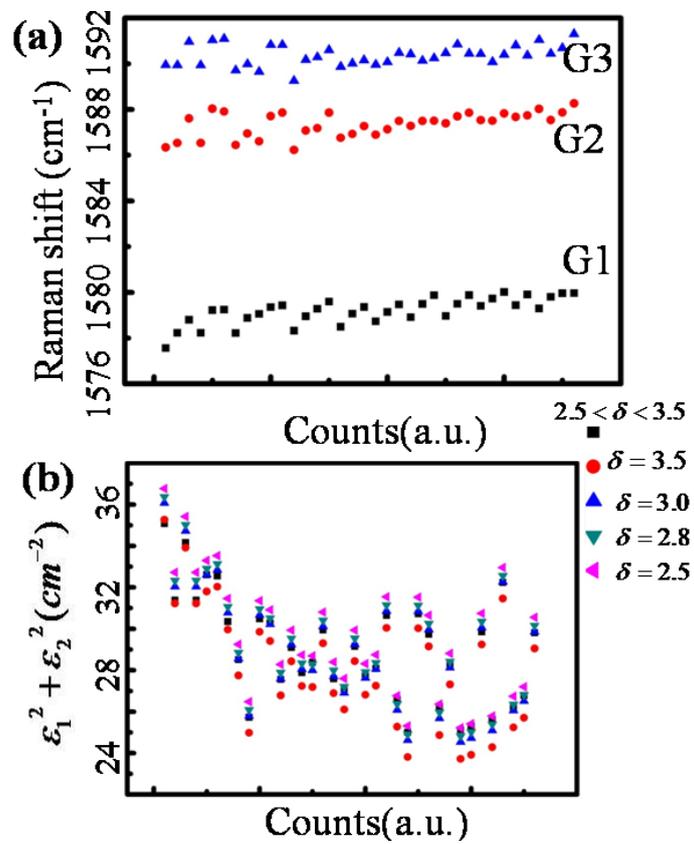

**Figure S1** Raman G bands of nanoripple on trilayer graphene, which is symmetric. The Symmetric G peak indicates that the strain is not the reason for the G peak splitting in trilayer graphene.

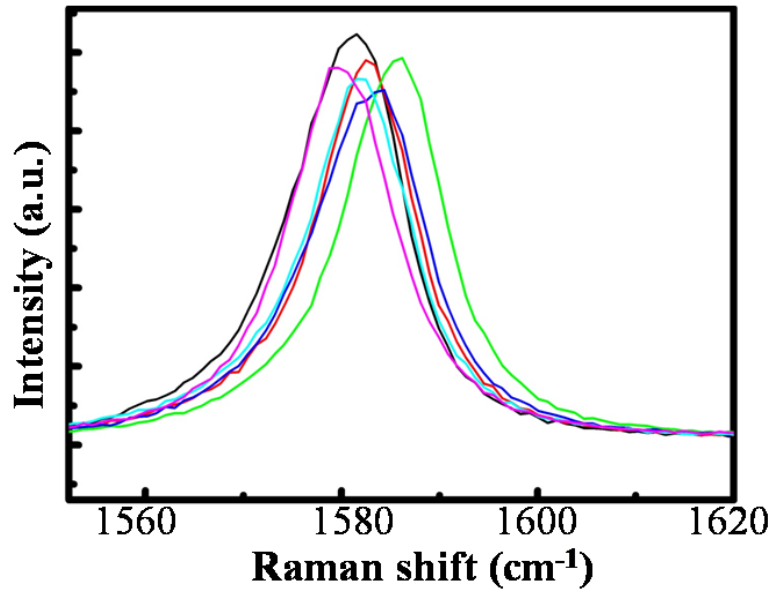

**Figure S2** (a) Variation of G1, G2 and G3 peak positions as a function of sample locations. (b) the variation of $\varepsilon_1^2 + \varepsilon_2^2$ at different value of $\delta$. $\varepsilon_1^2 + \varepsilon_2^2$ has only a slight dependence on the value of $\delta$ (2.5 to 2.8), however, varies from location to location in a broad range (9 to 30). This further confirms that the flucatution of G peak position is related with the varialbe $\varepsilon$ in the TLG, which should be related with spatially varied unintentional doping level at a microscopic level.

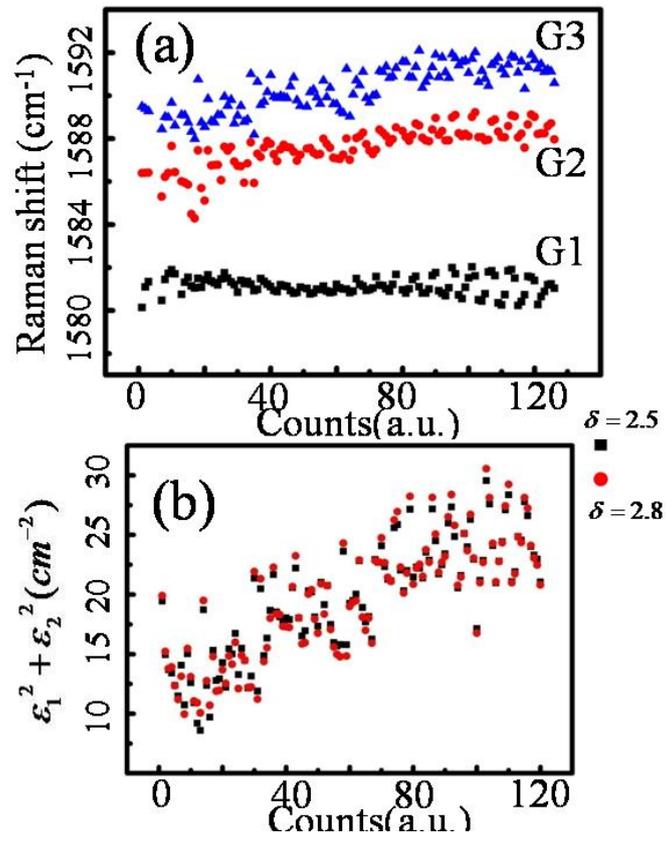